# STATISTICAL ANALYSIS OF AIR TRAFFIC IN LATVIAN REGION

Helen Afanasyeva
Riga Technical University
Kalku street No 1,
LV–1658 Riga, Latvia
E-mail: jelena_a@latnet.lv

**KEYWORDS**

**ABSTRACT**

The goal of the research is statistical analyzes of air traffic in airport "Riga" zone. Special statistical methods oriented to the concrete object area – airspace of Latvia are developed. Some experiments are made to discover season's and during twenty-four hours unstationarity of this process. Air traffic intensity for some stationary period for some airways is estimated.

**INTRODUCTION**

Airspace of Latvia on account of its geographical position takes on special significance for international aviation communication . It is most profitable for the Republic to use its airspace as intensively as possible , since it brings an evident income into the state budget . That is why new researchs in this field are of vital importance nowadays. Thereupon air traffic control of the airport 'Riga' carries on constant work aimed at the search of a more optimal structure and organization of using the airspace. On basis of researchs of this sort air routes over the territory under control are introduced and undergo necessary corrections.

One of the problems connected with air traffic control is that with the growing number of aircraft which are in the dispatcher's zone at the same time the difficulty of air traffic control increases, and at times it becomes problematic for the flying control officer to make an efficient decision. The fulfilment of most of the operations is connected with the shortage of time as a result of which the flying control officer's work requires considerable physical effort. To solve these and other problems it is necessary to investigate the streams of aircraft and build appropriate models for analysis and prognostication of different characteristics concerning the systems' functioning. Models built on the results of the statistical analysis may be used for the solution of the following main problems:

- the study of air traffic characteristics in concrete zones of control;
- the research of the influence of different factors (a human element, external and internal ones) on the efficiency of the functioning of air traffic control system;
- the determination of optional dimensions and structure of different districts and zones of air traffic control;
- the examination of the existing methods of control and working out of new ones under the raising of functional tasks of the system;
- the research of the loading and carrying (handling) capacity both of the system itself and its elements in particular;
- the prognostication of the efficiency of the existing structure of the airspace under the increasing air traffic intensity.

The study of these characteristics allows to determine low points of the systems, gives information for the improvement of the systems and designing of flying control officers' work.

The characters of the present-day systems of air traffic control are a high dynamism of its processes, the complexity of the object of control and the system itself, the shortage of time necessary for making certain decisions concerning the control, the necessity to solve non-typical problems, the rise of peak load. The statistical uncertainty of flying control officers' working prosses in air traffic control systems, a great number of various factors which influence its functioning, a great number of admission and outlet parameters require the accumulation of the material based on the experimental results, its thinking out, the estimation of the influence of different factors on the efficiency of the system. Working load which falls on flyng control officers ensuring the security of traffic and air situation control in a concrete zone is determined by the structure of airspace and the schemes of airways, hardware of navigation support of flights, communications facilities, control technology and organization of flights, the level of collection processes, identification and processing of data automation.

Statistical analysis of air traffic parameters such as air traffic intensity, aircraft distribution in echelons and airways, the ammount of aircraft that are under the control at the same time and so on, show that they are non-stationary in the course of time.

Thus, owing to heterogeneity and non-stationarity of the process of entrance of airplanes on certain air routes, it is non-advisable to use the available data at once for the selection of the appropriate distribution laws.

The information basis of the research is plan-reports of the flights, which were gained from the air-navigation system of collection and processing flight information database "SANS-2". They contain the following information concerning every transiting aircraft:
- the date of the flight's registration;
- the type of the aircraft;
- the identification code of the flight;
- the sending office;
- the point of destination;
- the point of entry into Riga FIR zone
- the point of outlet from Riga FIR zone
- the time of entry into Riga FIR zone

The following stages of preparation and analysis of the statistics may be singled out :
- the preparation of the data (statistics): confirming in its independence, after having uncovered its stationary periods, checking up the homogeneity of the data in the course of the periods;
- selecting the laws of distribution and its parameters.

After having considered the non-stationarity it is necessary to determine the distribution of time intervals (intensity) between the aircraft entering each air route during the pointed out stationary periods. If it is impossible to select the laws between the known distributions, then it is advisable to work out models of the mixed distributions estimation.

The aim of the work is statistical analysis of air traffic in the airport "Riga" zone with the help of modern statistical software package (MS Excel 2000 and Statistica 5.0).

As the result of it – stationary periods are revealed. After having considered non-stationarity it is necessary to determine the distribution of time intervals between the aircraft entering each air route during the pointed out stationary periods.

**NON-STATIONARITY ANALYSIS**

Thus, the first task is the quest of stationary periods. Non-stationarity may be:
- according to seasons
- according to days of the week
- according to the time of the day

**Season non-stationarity**

The first aim of the present work is to single out stationary periods according to months in the course of the year. According to the gathered statistics of different airlines, the following results which are represented in the form of a diagram in Figure 1 were gained.

We can see that the most inert period is observed in winter, then there is also no particular air traffic intensity in spring, where the number of flights increases only in May. Then in summer the activity increases and continues till the middle of autumn. After that a slump is observed, thus November and December are considered to be the most inert months.

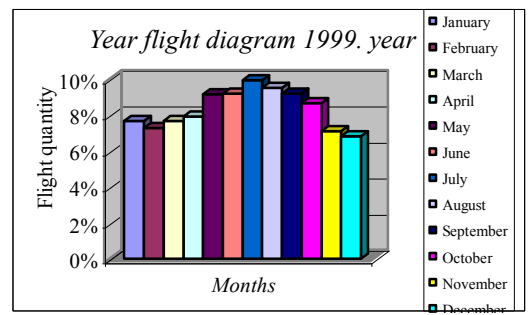

*Fig. 1 The Year flight diagram*

In this diagram you can see that the situation is approximately the same in January and February, March and April, which means approximately the same number of flights, and so we can regard this period as a stationary one. There is little difference in the level of intensity if we look at the following 2 months – May and June, so we consider this period to be stationary as well. Data analysis of everyday flights during the whole year shows that very often we can't find a stationary period longer than 2 months.

**Non-stationarity within 24 hours**

The next task is to determine non-stationarity according to days of the week, but here the situation was very unclear, as no particular dependance between the flights on the same days of the week was revealed.

The most interesting situation was noted in the changes of air traffic intensity within 24 hours. This can be seen in figure 2. Here the data for one of the air routes are presented. For each particular air route its own peculiar figure with different from each other stationary periods is observed.

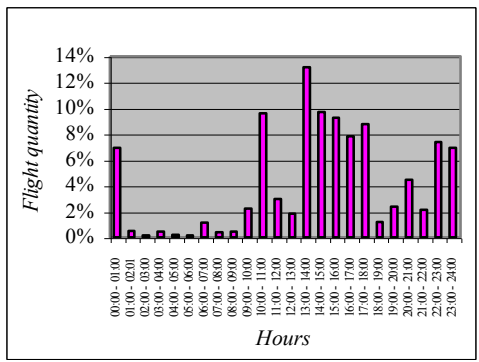

*Fig. 2 Flight's within 24 hours diagram (Ninta - Opoka).*

It can be distinctly seen how non-stationary this process is within 24 hours. During the day, especially in peak-hours from 2 p.m. till 6 p.m. a sudden tendency towards the increase of the number of flights is observed, whereas at night from 1 a.m till 8 a.m. the flight intensity is nearly zero.

As a result, stationary periods, different for each air route in particular within 24 hours, were revealed.

### THE DISTRIBUTION OF INTERVALS BETWEEN THE ARRIVALS OF AIRCRAFTS ANALYSIS

Now it is possible to determine the laws of interval distribution between the arrivals of aircraft for different air routes for each stationary period. An example of such a stationary analysis for one of the air routes is represented in Figure 3. Here a pronounced form of exponential distribution is seen. The hypothesis concerning this distribution was checked by means of the software package Statistica 5.0. The results of the check-up of the hypothesis for one of the cases were as follow: as Chi Square Test shows we can accept the hypothesis concerning the exponential distribution with the probability of 0.63. The check-up of hypothesis concerning other distributions is out of place here.

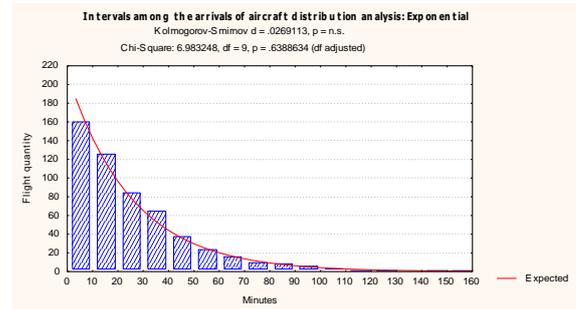

*Fig. 3 The distribution of intervals between the arrivals of aircraft analysis (Ninta – Opoka, stationary period 1 p. m. – 6 p. m.).*

The data analysis concerning other air routes was not so successful, though the exponential form still invariably existed to a certain extent. In theory this conclusion can be made on the assumption of the fact that at the entrance to a new air traffic control zone at one point of the entrance the streams from various directions join, and as is well known such a common stream is considered to be a Puasson one. In its turn it is well known that intervals between the accurrences in a Puasson stream are distributed according to the exponential law.

A situation when there is lack of data can arise. In this case we cannot cast aside (avoid) the consideration of the first and the last members of the period, as we did so far. It is necessary to use the technique of working with censored data.

Another similar problem arises in the presence of the mix of certain distributions (Figure 4). As a rule one of distributions in the mix is exponential.

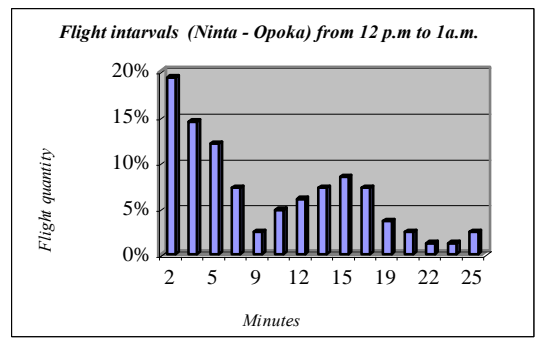

*Fig . 4 The mix of certain distributions example.*

The research of such laws of distribution is not an easy statistical problem. In this case we can't rely on packages of statistics where distributions are chosen between several the most popular kinds.

I'll give an example of the estimation of the law of the distribution of intervals between the arrivals of aircraft for the same air route Ninta-Opoka, but during another period (Figure 4). A mix of 2 distributions is used, and its general form is as follow

$$f(x) = p \cdot f_1(x) + (1-p) \cdot f_2(x), \quad \text{where}$$

$f_i(x)$    the unknown function of distribution density, i=1,2.

p -    the unknown parameter 0<p<1, probabilty of choosing a certain function of density.

As can be seen in the figure, it is most likely a mix of an exponential and normal distributions.

$$f_1(x) = \lambda \cdot x \cdot e^{-\lambda \cdot x}$$
$$f_2(x) = \frac{1}{\sqrt{2\pi}\sigma} e^{-\frac{1}{2}\left(\frac{x-\mu}{\sigma}\right)^2}$$

It is necessary that the unknown parameters of the obtained distribution: p, $\lambda$, $\mu$, $\sigma$.

There are 2 traditional methods of estimation of the distribution parameters (Rao, 1973):
1. The method of moments
2. The method of the maximum (highest possible) credibility

We'll use the method of moments.

On basis of the equalization of the calculated empiric moments $\mu_1^*, \mu_2^*, \mu_3^*, \mu_4^*$ and corresponding theoretical moments $\mu_1, \mu_2, \mu_3, \mu_4$ the following combined equations can be made:

$$\mu_\nu^* = \mu_\nu(p, \lambda, \mu, \sigma), \quad \nu = 1,2,3,4$$

namely,

$$\begin{cases} \mu_1^* = p\frac{1}{\lambda} + (1-p)\mu \\ \mu_2^* = p\frac{2}{\lambda^2} + (1-p)(\sigma^2 + \mu^2) \\ \mu_3^* = p\frac{6}{\lambda^3} + (1-p)(3\mu\sigma^2 + \mu^3) \\ \mu_4^* = p\frac{24}{\lambda^4} + (1-p)(3\sigma^4 + 6\mu^2\sigma^2 + \mu^4) \end{cases}$$

This combined equations can be solved by means of the numerical computing, used for the estimation of the unknown parameters of this distributions mix.

**CONCLUSIONS**

As a result of the carried out work the following main tasks were accomplished.

1. The non-stationarity of processes was analysed
   - Seasons
   - During 24 hours

2. On basis of the previous item air traffic intensity in different air routes was estimated, and so were the laws of interval distribution between the arrivals of aircraft

The results of the carried out statistical analysis were used for the construction of different imitating models of airspace.

It is planned to use the models for the construction of an imitating model of the intersection of air routes for the purpose of the probable dangerous approaches estimation.

**REFERENCES**

Andrews, D.F.; and Herzberg, A.M. 1985. *Data: a Collection of Problems from Many Fields for Student and Research Worker*. Springer, New York.

Rao, C.R. 1973. *Linear statistical inference and its applications*, 2nd edn. Wiley, New York.

Adams, A.; Bloomfield, D.; Booth, Ph.; England, P. *Investment Mathematics and Statistics*. Graham&Trotman,London.